\documentclass[10pt]{iopart}
\usepackage{graphicx}

\begin{document}

\title[Relaxation time in a non-conserving system]{Relaxation time in a non-conserving driven-diffusive system with parallel dynamics}

\author{S. R. Masharian $^1$,\;\;F. H. Jafarpour$^{2}$ and \\ A. Aghamohammadi $^{2}$}
\address{$^1$Islamic Azad University, Hamedan Branch, Hamedan, Iran}
\address{$^{2}$Physics Department, Bu-Ali Sina University, 65174-4161 Hamedan, Iran} 
\begin{abstract}
We introduce a two-state non-conserving driven-diffusive system in one-dimension under a discrete-time updating scheme. We show that the steady-state of the system can be obtained using a matrix product approach. On the other hand, the steady-state of the system can be expressed in terms of a linear superposition Bernoulli shock measures with random walk dynamics. The dynamics of a shock position is studied in detail. The spectrum of the transfer matrix and the relaxation times to the steady-state have also been studied in the large-system-size limit.
\end{abstract}
\maketitle
\section{Introduction}
Exactly solvable systems have been of great interests to physicists for years. Among these systems there are one-dimensional non-equilibrium systems which have unique critical and collective properties \cite{SMG}-\cite{TML}. Some of these properties, which usually can not be found in their equilibrium counterparts, are out-of-equilibrium phase transitions and shock formations. Despite of these interesting properties, the number of one-dimensional driven-diffusive systems belonging to the family of non-equilibrium systems, which can be solved analytically is very limited.\\
Over the past two or three decades, the one-dimensional driven-diffusive systems have been studied extensively from different point of views. It has been shown that some of these systems can be mapped onto the zero-range processes \cite{EH}. Some of them, on the other hand, are related to the lattice-path 
models \cite{BJJK,BDR,FJH}. It is known that in some of these systems the dynamics of a shock (or equivalently a sharp discontinuity in the density of particles in the system) is similar to that of a simple random walker. In this way the study of a system with large number of degrees of freedom reduces to the study of a system with fewer number of degrees of freedom \cite{KJS}.\\
It has been shown that in the continuous-time updating scheme there are only three two-state driven-diffusive systems with open boundaries and nearest-neighbor interactions in which a single shock can evolve in time with a random walk dynamics \cite{KJS}. The steady-state of these systems can be written as matrix product states and that their associated quadratic algebras have two-dimensional matrix representations \cite{JM1}. This idea has been also generalized to the systems containing multiple species of particles with next-nearest-neighbor interactions \cite{RS,PS,JM2}. However, in the discrete-time updating scheme there is no such a classification and only a couple of examples exist \cite{PIS,BS}.\\
In this paper we introduce a two-states one-dimensional driven-diffusive system with open boundaries. The updating scheme is parallel which consists of two steps. This scheme is sometimes called the sublattice-parallel updating scheme. This system does not belong to that three-member family of two-states driven-diffusive systems with open boundaries and  nearest-neighbor interactions. The steady-state of this system can be obtained using a matrix product approach (for a review see \cite{BE}). On the other hand, the steady-state of the system can be written as a superposition of shocks with simple random walk dynamics. The time evolution of the probability distribution function can also be studied in special cases. The eigenvalues and eigenvectors of the time evolution operator can be obtained using a simple plane wave ansatz. The relaxation times can be calculated exactly for large-system-size.\\
This paper is organized as follows: in the first section we define the model. By introducing two product shock measures we show that the time evolution of the shock positions are similar to those of two simple one-dimensional random walkers in discrete time. We build the steady-state of the system as a linear superposition of these shocks. In the second section we review the basics of the matrix product approach and show that the steady-state of the system can also be found using this approach. In the third section we investigate the relaxation to the steady-state. In this direction we try to find the largest eigenvalue of the time evolution operator using the standard plane wave ansatz by imposing some restrictions on the microscopic reaction probabilities. We will finally discuss about the possible extensions to this model and open questions. 
\section{Definition of the process}
We consider a lattice of length $2L$ with open boundaries. Each lattice site can be either occupied by a particle $A$ or a vacancy $\emptyset$. The particles are injected and extracted from both ends of the lattice with certain probabilities. At left boundary (the first lattice site) we have:
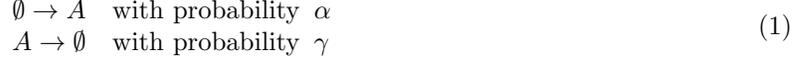
\begin{equation}
\label{Rules1}
\begin{array}{ll}
\emptyset \rightarrow A & \mbox{with probability} \; \; \alpha \\
A \rightarrow \emptyset& \mbox{with probability} \; \; \gamma
\end{array}
\end{equation}
while at right boundary (the last lattice site):
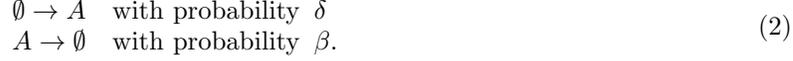
\begin{equation}
\label{Rules2}
\begin{array}{ll}
 \emptyset\rightarrow A & \mbox{with probability} \; \; \delta \\
A\rightarrow \emptyset & \mbox{with probability} \; \; \beta.
\end{array}
\end{equation}
We assume that in the bulk of the lattice the particles hop to the left and right while interacting with each other. Apart from diffusion we have pair annihilation and coagulation of particles to the left and right. The reaction rules are as follows:
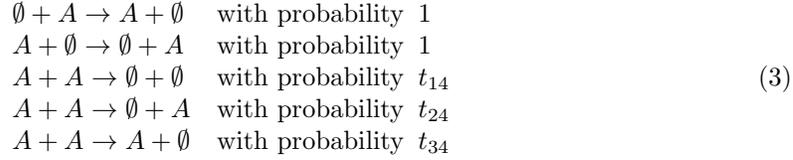
\begin{equation}
\label{Rules3}
\begin{array}{ll}
\emptyset+A \rightarrow A+\emptyset & \mbox{with probability} \; \; 1 \\
A+\emptyset \rightarrow \emptyset+A & \mbox{with probability} \; \; 1 \\
A+A \rightarrow \emptyset+\emptyset & \mbox{with probability} \; \; t_{14} \\
A+A \rightarrow \emptyset+A & \mbox{with probability} \; \; t_{24} \\
A+A \rightarrow A+\emptyset & \mbox{with probability} \; \; t_{34}
\end{array}
\end{equation}
under the constraint:
\begin{equation}
\label{Constraint}
t_{14}+t_{24}+t_{34}=1
\end{equation} 
which will be discussed in the next section. The time is discrete  and that the updating scheme applied is called the sublattice-parallel which is defined as follows: we divide the bulk dynamics into two half-time steps. In the first half-time step the pairs of neighboring sites ($2k,2k+1$) for $k=1,\cdots,L-1$ and also the first and the last lattice sites are updated. In the second half-time step the pairs of neighboring sites ($2k-1,2k$) for $k=1,\cdots,L$ are updated.\\
The time evolution of the probability distribution vector $\vert P(t) \rangle$ is given by the following master equation:
\begin{equation}
\label{TE}
T \vert P(t) \rangle
= \vert P(t+1) \rangle.
\end{equation}
where $T$ in (\ref{TE}) is called the transfer matrix. The transfer matrix $T$ is given by a product of two factors $T_1$ and $T_2$ as $T=T_1T_2$. These factors are defined as follows:
$$
\begin{array}{ll}
T_1 = {\cal L} \otimes {\cal T} \otimes \ldots \otimes {\cal T} \otimes {\cal R} \,\;=\;\,{\cal L} \otimes {\cal T}^{\otimes (L-1)} \otimes {\cal R} \\ \\
T_2 =  {\cal T} \otimes {\cal T} \otimes \ldots \otimes{\cal T}  \,\;=\;\,  {\cal T}^{\otimes L} 
\end{array}
$$
where ${\cal T}$, ${\cal L}$ and ${\cal R}$ are given by:
\begin{equation}
\fl
\label{TM}
{\cal T}=\left(
\begin{array}{cccc}
1&0&0&t_{14}\\
0&0&1&t_{24}\\
0&1&0&t_{34}\\
0&0&0&0\\
\end{array}
\right),
\cal{L}=\left(
\begin{array}{cc}
1-\alpha&\gamma\\
\alpha&1-\gamma\\
\end{array}
\right),
\cal{R}=\left(
\begin{array}{cc}
1-\delta&\beta\\
\delta&1-\beta\\
\end{array}
\right).
\end{equation}
The matrix $\cal T$ is written in the basis $(\emptyset\emptyset,\emptyset A,A \emptyset,AA)$. The matrices $\cal L$ and $\cal R$ are also written in the basis $(\emptyset,A)$. In the long-time limit the system approaches its steady- state and one has:
\begin{equation}
\label{SS}
T \vert P^{*} \rangle = \vert P^{*} \rangle.
\end{equation}
In the following sections we will investigate the steady-state of the system using two different approaches.
\section{Steady-state as a superposition of shocks}
In this section we show that one can construct the steady-state of the system as a linear superposition of product shock measures. At the outset we define two product shock measures and investigate their time evolution under the transfer matrix $T$ defined in the previous section. The product shock measures at even sites $2k$ ($k=1,\cdots,L$) will be denoted by $\vert 2k\rangle$ and the product shock measures at odd sites $2k+1$ ($k=0,\cdots,L$) will be denoted by $\vert 2k+1 \rangle$. Defining the notation:
$$
\vert \rho_{1,2}^{o,e}\rangle=\left(\begin{array}{c}
1-\rho_{1,2}^{o,e} \\ \rho_{1,2}^{o,e}
\end{array}\right)
$$
we write:
\numparts
\begin{eqnarray}
\label{SM1}
\vert 2k \rangle=(\vert \rho_{1}^{o}\rangle\otimes\vert \rho_{1}^{e}\rangle)^{\otimes{k-1}}\otimes
\vert \rho_{1}^{o}\rangle\otimes\underbrace{\vert \rho_{2}^{e}\rangle}_{2k}\otimes(\vert \rho_{2}^{o}\rangle\otimes\vert \rho_{2}^{e}\rangle)^{\otimes{L-k}} \\ 
\label{SM2}
\vert 2k+1 \rangle=  (\vert \rho_{1}^{o}\rangle\otimes\vert \rho_{1}^{e}\rangle)^{\otimes{k}} \otimes
\underbrace{ \vert \rho_{2}^{o} \rangle}_{2k+1}\otimes \vert \rho_{2}^{e}\rangle \otimes
(\vert \rho_{2}^{o}\rangle\otimes\vert \rho_{2}^{e}\rangle)^{\otimes{L-k-1}}.
\end{eqnarray}
\endnumparts
Note that the lattice site $2L+1$ is an auxiliary lattice site so that the product shock measure $\vert 2L+1\rangle$ indicates a distribution of particles with densities $\rho_{1}^{o}$ and $\rho_{1}^{e}$ at odd and even lattice sites respectively. In this case the shock front can be considered to be between the lattice sites $2L$ and $2L+1$. We now investigate the time evolutions of (\ref{SM1}) and (\ref{SM2}) under the transfer matrix $T$. Let us consider the following values for the density of particles:
\begin{equation}
\label{Density}
\rho_1^{o}=0\;\;,\;\;\rho_2^{o}=\delta\;\;,\;\;\rho_1^{e}=\alpha\;\;,\;\;\rho_2^{e}=0.
\end{equation}
These values beside the constraint (\ref{Constraint}) provide the necessary and sufficient conditions which guarantee that the shock position has a simple random walk dynamics. After some calculations we find:
\begin{equation}
\begin{array}{l}
\label{TEE}
\fl T\vert 1 \rangle=\pi_{r1}  \vert 2 \rangle+\pi_{r2} \vert 3 \rangle+(1-\pi_{r1}-\pi_{r2}) \vert1 \rangle\\
\fl T\vert 2k \rangle=\pi_{l} \vert 2k-1 \rangle+\pi_{r} \vert 2k+1 \rangle+\pi_{s} \vert 2k \rangle \;\;\mbox{for} \;\;1 \leq k \leq L\\ 
\fl T\vert 2k+1\rangle=\pi_{l} \pi_{s} \vert 2k \rangle +\pi_{r} \pi_{s} \vert 2k+2 \rangle + \pi_{r}^{2} \vert 2k+3 \rangle+\pi_{l}^2\vert 2k-1 \rangle \nonumber\\
                               +(\pi_{s}+2\pi_{l}\pi_{r})\vert 2k+1 \rangle\;\;\mbox{for} \;\; 1 \leq k \leq L-1      \\
 \fl T\vert 2L+1\rangle=\pi_{l1} \vert 2L \rangle+\pi_{l2} \vert 2L-1 \rangle+(1-\pi_{l1}-\pi_{l2}) \vert 2L+1 \rangle                       
\end{array}
\end{equation}  
in which we have defined $\pi_{s}=1-\pi_{r}-\pi_{l}$ and that:
\begin{equation}
\label{CON}
\begin{array}{l}
\pi_{r}=1-\delta+\delta t_{24}\;\;,\;\;\pi_{l}=1-\alpha+\alpha \; t_{34}\\
\pi_{r1}=\frac{\alpha(1-\delta)+(1-\gamma)\delta}{\alpha} \pi_{s}\;\;,\;\;\pi_{r2}=\frac{\alpha(1-\delta)+(1-\gamma)\delta}{\alpha} \pi_{r}\\
\pi_{l1}=\frac{\delta(1-\alpha)+(1-\beta)\alpha}{\delta} \pi_{s}\;\;,\;\;\pi_{l2}=\frac{\delta(1-\alpha)+(1-\beta)\alpha}{\delta} \pi_{l}.
\end{array}
\end{equation}
The equations (\ref{TEE}) indicate that the shock position evolves in time with a dynamics which is similar to that of a simple random walker and when it reaches to the boundaries of the lattice it reflects from them.\\
Using the equations (\ref{TEE}) it is now easy to construct the steady-state of the system defined in (\ref{Rules1}-\ref{Rules3}). Let us consider a linear superposition of the above defined shocks as follows:
\begin{equation}
\label{SPS}
\vert P^{\ast} \rangle = \frac{1}{Z_{L}}\sum_{k=1}^{2L+1}P^{\ast}_{k} \vert k \rangle
\end{equation}
in which $Z_{L}$ is a normalization factor given by:
\begin{equation}
\label{NF}
Z_L=\sum_{k=1}^{2L+1}P^{\ast}_{k}
\end{equation}
If $\vert P^{\ast} \rangle$ is the steady-state then it should satisfy (\ref{SS}). It turns out that the coefficients $P^{\ast}_k$'s are given by:
\begin{equation}
\label{SSP}
\begin{array}{l}
P^{\ast}_{2k}=\pi_{s}(\frac{\pi_{r}}{\pi_{l}})^{2k} \; \mbox{for} \; k=2,\cdots,L-1\\ 
P^{\ast}_{2k+1}=(\frac{\pi_{r}}{\pi_{l}})^{2k+1} \; \mbox{for} \; k=1,\cdots,L-1\\ 
P^{\ast}_{1}=\frac{\pi_{r}^2}{\pi_{r1}\pi_{r}+\pi_{r2}(1-\pi_{s})}\;(\frac{\pi_{r}}{\pi_{l}})\\ 
P^{\ast}_{2}=\frac{\pi_{r}(\pi_{r2}\; \pi_{s}+\pi_{r1}(1-\pi_{r}))}{\pi_{r1}\pi_{r}+\pi_{r2}(1-\pi_{s})} \; (\frac{\pi_{r}}{\pi_{l}})^2\\ 
P^{\ast}_{2L}=\frac{\pi_{l}(\pi_{l2}\; \pi_{s}+\pi_{l1}(1-\pi_{l}))}{\pi_{l1}\pi_{l}+\pi_{l2}(1-\pi_{s})} \; (\frac{\pi_{r}}{\pi_{l}})^{2L}\\ 
P^{\ast}_{2L+1}=\frac{\pi_{l}^2}{\pi_{l1}\pi_{l}+\pi_{l2}(1-\pi_{s})}\;(\frac{\pi_{r}}{\pi_{l}})^{2L+1}.
\end{array}
\end{equation}
Note that if $\alpha=\delta=0$ then an empty lattice is the steady-state of the system which can be written as:
\begin{equation}
\label{SS1}
\vert P^{\ast} \rangle=\left(
\begin{array}{c}
1\\
0\\
\end{array}
\right)^{\otimes 2L}.
\end{equation}
On the other hand, if $\alpha=\beta=1$ and $0 \le \delta,\gamma \le 1$ the steady-state of the system is given by:
\begin{equation}
\label{SS2}
\vert P^{\ast} \rangle=[\left(
\begin{array}{c}
1\\
0\\
\end{array}
\right)\otimes
\left(
\begin{array}{c}
0\\
1\\
\end{array}
\right)]^{\otimes L}.
\end{equation}
Finally, if $\delta=\gamma=1$ and $0 \le \alpha,\beta \le 1$ the steady-state of the system is given by:
\begin{equation}
\label{SS3}
\vert P^{\ast} \rangle=[\left(
\begin{array}{c}
0\\
1\\
\end{array}
\right)\otimes
\left(
\begin{array}{c}
1\\
0\\
\end{array}
\right)]^{\otimes L}.
\end{equation}
In the next section we describe how one can obtain the steady-state of the system using a matrix product approach.
\section{Matrix product approach}
According to the matrix product approach the steady-state of the system $\vert P^{\ast} \rangle$ can be written as \cite{HH}:
\begin{equation}
\label{MPSS}
\vert P^{\ast}\rangle= \frac{1}{Z_L} \langle \langle W \vert
\left[\left( \begin{array}{c}
{\hat E} \\ {\hat D} \end{array} \right) \otimes \left( \begin{array}{c}
E \\ D \end{array} \right)\right]^{\otimes L}
\vert V \rangle \rangle
\end{equation}
in which the operators $\hat E$ and $\hat D$ ($E$ and $D$ ) stand for the presence of a vacancy and a particle at an odd (even) lattice site respectively. 
The denominator $Z_L$ is a normalization factor. The four operators $\hat{E}$, $\hat{D}$, $E$, and $D$ besides the vectors $\vert V \rangle\rangle$ and
$\langle \langle W \vert$ satisfy a quadratic algebra which results from the following relations:
\begin{eqnarray}
\label{ALG}
&&{\cal T} \, \left[ \Bigl(
    \begin{array}{c} E \\ D
    \end{array}\Bigr) \otimes \Bigl(
    \begin{array}{c} \hat{E} \\ \hat{D}
    \end{array} \Bigr) \right] \;=\; \Bigl(
    \begin{array}{c} \hat{E} \\ \hat{D}
    \end{array} \Bigr) \otimes \Bigl(
    \begin{array}{c} E \\ D
    \end{array}\Bigr),  \nonumber\\
&&\langle\langle W \vert {\cal L} \Bigl(
    \begin{array}{c} \hat{E} \\ \hat{D}
    \end{array} \Bigr) \;=\; \langle \langle W \vert \Bigl(
    \begin{array}{c} E \\ D
    \end{array}\Bigr),\\
&&{\cal R} \Bigl(
    \begin{array}{c} E \\ D
    \end{array}\Bigr) \vert V \rangle\rangle\;=\; \Bigl(
    \begin{array}{c} \hat{E} \\ \hat{D}
    \end{array} \Bigr) \vert V \rangle\rangle  \nonumber.
\end{eqnarray}
The normalization factor $Z_L$ is also given by:
\begin{equation}
Z_L=\langle \langle W \vert (C\hat{C})^L \vert V\rangle \rangle
\end{equation}
in which $C=D+E$ and $\hat{C}=\hat{D}+\hat{E}$. Assuming $\hat{C}=C$ by using (\ref{TM}) and (\ref{ALG}) we find the following quadratic algebra associated with the process (\ref{Rules1})-(\ref{Rules3}):
\begin{equation}
\label{BulkAlgebra}
\begin{array}{l}
 \hat{D}D = 0 \; ,
[\hat{D}, C] = (t_{34}-1)D \hat{D}\; , \;[C,D]=(t_{24}-1)D \hat{D},\\
\langle\langle W \vert (\alpha C+(1-\alpha-\gamma)\hat{D}-D)=0,\\
(\delta C+(1-\beta-\delta)D-\hat{D})\vert V \rangle\rangle=0.\\
\end{array}
\end{equation}
The quadratic algebra (\ref{BulkAlgebra}) has a two-dimensional matrix representation provided that the constraints (\ref{Constraint}) and (\ref{Density}) are fulfilled. The representation is then given by :
\begin{equation}
\begin{array}{l}
\label{BulkRep}
\hat{D} = \left( \begin{array}{cccc}
\rho_{2}^{o}& & & 0 \\
\hat{d} & & & \frac{\pi_{r}}{\pi_{l}}\rho_{1}^{o}
\end{array} \right) \; , \;
\hat{E} = \left( \begin{array}{cccc}
1-\rho_{2}^{o}& & & 0 \\
-\hat{d} & & & \frac{\pi_{r}}{\pi_{l}}(1-\rho_{1}^{o})
\end{array} \right) \; ,\; \\ \\
D = \left( \begin{array}{cccc}
\rho_{2}^{e}  & & & 0 \\
d & & & \frac{\pi_{r}}{\pi_{l}}\rho_{1}^{e}
\end{array} \right) \; , \;
E = \left( \begin{array}{cccc}
1-\rho_{2}^{e} & & & 0  \\
-d & & & \frac{\pi_{r}}{\pi_{l}}(1-\rho_{1}^{e})
\end{array} \right) \; , \; \\ \\
\langle\langle W \vert = (1, \frac{\alpha-\delta(\alpha+\gamma-1)}{\hat{d}(\alpha+\gamma-1)+d}) \; , \;
\vert V \rangle \rangle= \left( \begin{array}{c}
1 \\ \frac{-\pi_{l}(\hat{d}+d(\delta +\beta-1))}{\pi_{r}(\alpha(\delta+\beta-1)-\delta)} \end{array} \right)
\end{array}
\end{equation}
where $d=-\hat{d}(\frac{1-t_{24}}{1-t_{34}})$. Note that the algebra (\ref{BulkAlgebra}) has also one-dimensional representations which are associated with the special cases discussed at the end of the previous section. In this case the operators $\hat E$, $\hat D$, $E$ and $D$ are in general complex numbers. For $\alpha=\delta=0$ one finds $\hat{D}=D=0$ and $C=E=\hat{E}$. On the other hand, for $\alpha=\beta=1$ and $0 \le \delta,\gamma \le 1$ one finds $\hat{D}=E=0$ and $C=\hat{E}=D$. Finally, for $\delta=\gamma=1$ and $0 \le \alpha,\beta \le 1$ one finds $D=\hat{E}=0$ and $C=\hat {D}=E$. Apart from these special cases, the two-dimensional matrix representation given above is the exact result. Since the steady-state of the system is unique, it is not difficult to show that the steady-state of the system in terms of a linear superposition of shocks given in (\ref{SPS}) is exactly equal to one given in (\ref{MPSS}) in terms of a product of non-commuting operators.
\section{Relaxation to the steady-state}
In the following we study the relaxation time to stationarity in the random walk picture using a plane wave ansatz. Let us define $P_{k}(t)$ as the probability of finding the shock position at time $t$ at the lattice site $k$. As $t \rightarrow \infty$ the probability $P_{k}(t)$ converges to $P^{\ast}_{k}$ given in (\ref{SSP}). We write:
\begin{equation}
\label{TDPV}
\vert P(t) \rangle=\sum_{k=1}^{2 L+1} P_{k}(t) \vert k \rangle.
\end{equation}
The time evolution of (\ref{TDPV}) is governed by the master equation (\ref{TE}). Using (\ref{TE}), (\ref{TEE}) and (\ref{TDPV}) and after some calculations we find:
\begin{equation}
\label{TDP}
\begin{array}{l}
\fl P_{1}(t+1)=\pi_{l} P_{2}(t)+\pi_{l}^{2} P_{3}(t)+(1-\pi_{r1}-\pi_{r2})  P_{1}(t)\\ 
\fl P_{2}(t+1)=\pi_{r1} P_{1}(t)+\pi_{l}\pi_{s} P_{3}(t)+\pi_{s}  P_{2}(t)\\ 
\fl P_{3}(t+1)=\pi_{r2} P_{1}(t)+\pi_{r}P_{2}(t)+\pi_{l}  P_{4}(t)+\pi_{l}^2 P_{5}(t)+(\pi_{s}+2 \pi_{l}\pi_{r} )P_{3}(t)\\ 
\fl P_{2k}(t+1)=\pi_{r} \pi_{s} P_{2k-1}(t)+\pi_{l} \pi_{s}P_{2k+1}(t)+\pi_{s}  P_{2k}(t) \;\;\;\mbox{for} \;\;\; 2 \leq k \leq L-1\\ 
\fl P_{2k+1}(t+1)=\pi_{r}^{2} P_{2k-1}(t)+\pi_{r}P_{2k}(t)+\pi_{l}  P_{2k+2}(t)+\pi_{l}^2 P_{2k+3}(t)+\\
(\pi_{s}+2 \pi_{l}\pi_{r} )P_{2k+1}(t)\;\;\;\mbox{for} \;\;\; 2 \leq k \leq L-2\\ 
\fl P_{2L-1}(t+1)=\pi_{r}^{2} P_{2L-3}(t)+\pi_{r}P_{2L-2}(t)+\pi_{l}  P_{2L}(t)+\pi_{l2} P_{2L+1}(t)+\\
(\pi_{s}+2 \pi_{l}\pi_{r} )P_{2L-1}(t)\\ 
\fl P_{2L}(t+1)=\pi_{r} \pi_{s} P_{2L-1}(t)+\pi_{l1} P_{2L+1}(t)+\pi_{s}  P_{2L}(t)\\ 
\fl P_{2L+1}(t+1)=\pi_{r}P_{2L}(t)+\pi_{r}^{2}P_{2L-1}(t)+(1-\pi_{l1}-\pi_{l2})P_{2L+1}(t)\\ 
\end{array}
\end{equation}
Before going any farther let us have a note of the drift velocity $v$ and diffusion coefficient $D$ of the shock front in an infinite system. Considering the bulk equations in  (\ref{TDP}) for an infinite system, one can calculate $v$ and $D$ according to the approach used in \cite{PIS}. The moments are given by:
\begin{equation}
\langle k^{n} (t) \rangle=\sum_{k=-\infty}^{\infty}k^{n}P_{k}(t)=\langle k^{n}_{e}(t)\rangle+\langle k^{n}_{o}(t)\rangle
\end{equation}
by defining:
$$
\begin{array}{l}
\langle k^{n}_{e}(t)\rangle:=\sum_{k=-\infty}^{\infty}(2k)^{n}P_{2k}(t) \\ \\
\langle k^{n}_{o}(t)\rangle:=\sum_{k=-\infty}^{\infty}(2k+1)^{n}P_{2k+1}(t).
\end{array}
$$
Using the definitions:
$$
\begin{array}{l}
v=\lim_{t\rightarrow \infty}(\langle k(t+1)\rangle-\langle k(t)\rangle)\\ \\
D=\lim_{t\rightarrow \infty}((\langle k^{2}(t+1)\rangle-\langle k(t+1)\rangle^{2})-(\langle k^{2}(t)\rangle-\langle k(t)\rangle^{2})).
\end{array}
$$
one finds:
\begin{equation}
v=2\frac{\pi_{r}-\pi_{l}}{1+\pi_{s}}\;\;,\;\;D=2\frac{1-\pi_{s}}{1+\pi_{s}}(1-{\frac{v}{2}}^2).
\end{equation}
As can be seen these expressions are exactly the same as those belong to the asymmetric simple exclusion process calculated in \cite{PIS}. This means that in the long-time limit these properties do not depend on the microscopic dynamics.\\
Finding a general solution for the equations (\ref{TDP}) is a formidable task; therefor, in what follows we consider the case $\pi_{s}=0$ which results in  $\pi_{r1} =\pi_{l1}=0$. In this case only the shocks at odd lattice sites appear in (\ref{TDPV}). Simplifying the equations (\ref{TDP}) results in:
\begin{equation}
\label{RME}
\begin{array}{ll}
\fl P_{1}(t+1)=\pi_{l}^{2} P_{3}(t)+(1-\pi_{r2})  P_{1}(t)\\ 
\fl P_{3}(t+1)=\pi_{r2} P_{1}(t)+\pi_{l}^2 P_{5}(t)+2 \pi_{l}\pi_{r} P_{3}(t)\\ 
\fl P_{2k+1}(t+1)=\pi_{r}^{2} P_{2k-1}(t)+\pi_{l}^2 P_{2k+3}(t)+2 \pi_{l}\pi_{r} P_{2k+1}(t)\;\;\; \mbox{for} \;\;\; 2 \leq k \leq L-2\\ 
\fl P_{2L-1}(t+1)=\pi_{r}^{2} P_{2L-3}(t)+\pi_{l2} P_{2L+1}(t)+2 \pi_{l}\pi_{r} P_{2L-1}(t)\\ 
\fl P_{2L+1}(t+1)=\pi_{r}^{2}P_{2L-1}(t)+(1-\pi_{l2})P_{2L+1}(t)
\end{array}
\end{equation}
where $\pi_{l}+\pi_{r}=1$. For the eigenfunction we suggest:
\begin{equation}
\label{EIGEN}
P_{k}(z,t)=\Lambda^{t}(z)\tilde{P}_{k}(z) \;\;\;\mbox{for odd}\;\;\;k
\end{equation}
in which $\Lambda(z)$ is the eigenvalue. By considering:
\begin{equation}
\label{PROP}
\fl
\begin{array}{l}
\tilde{P}_{2k+1}(z)=(\frac{\pi_{r}}{\pi_{l}})^{\frac{2k+1}{2}} (a(z) z^{2k+1} - a(z^{-1}) z^{-2k-1}) \; \; \mbox{for} \; 1 \leq k \leq L-1 \\ \\
\tilde{P}_{1}(z)=\tilde{C}_1 \; (\frac{\pi_{r}}{\pi_{l}})^{\frac{1}{2}} (a(z) z- a(z^{-1}) z^{-1})\\ \\
\tilde{P}_{2L+1}(z)=\tilde{C}_2 \; (\frac{\pi_{r}}{\pi_{l}})^{\frac{2L+1}{2}} (a(z) z^{2L+1} - a (z^{-1}) z^{-2L-1})\\ 
\end{array}
\end{equation}
and replacing them in (\ref{RME}) and after some straightforward calculations one finds:
$$
a(z)=(\frac{\pi_{r}}{\pi_{l}})^{\frac{1}{2}}z^{-1}(\pi_{r} \pi_{l}z^{-2}+(1-\pi_{r2}-\Lambda(z))\tilde{C}_1)\;,\;
\tilde{C}_1=\frac{\pi_{r}^2}{\pi_{r2}}\;,\;\tilde{C}_2=\frac{\pi_{l}^2}{\pi_{l2}}
$$
and the eigenvalues:
\begin{equation}
\Lambda(z)=\pi_{r} \pi_{l}(z+z^{-1})^{2}.
\end{equation}
It turns out that the proper $z$'s belong to two groups: The $z$'s of the first group contains four $z$'s which are $z=\pm (\frac{\pi_{r}}{\pi_{l}})^{\pm{\frac{1}{2}}}$ resulting in $\Lambda=1$ and therefore correspond to the steady-state.  The $z$'s of the second group satisfy the following equation:
\begin{equation}
\label{ZE}
z^{4L}=\frac{(1+A z^2)(1+Bz^2)}{(z^2+A)(z^2+B)}
\end{equation}
where $A=\frac{\pi_{l2}-\pi_{l}^2}{\pi_{l}\pi_{r}}$ and $B=\frac{\pi_{r2}-\pi_{r}^2}{\pi_{l}\pi_{r}}$. The equation (\ref{ZE}) has  $4L+4$ roots; however, the roots $z=\pm 1$ and $z=\pm \rmi$ result in a zero eigenfunction and have to be excluded. One also notes that since $\Lambda(z)=\Lambda(-z)$ and also $\Lambda(z)=\Lambda(\frac{1}{z})$ the remaining $4L$ roots result in $L$ eigenvalues. \\
The roots of a very similar equation to (\ref{ZE}) has already been investigated in \cite{KA}. Finding the roots of (\ref{ZE}) for an arbitrary $L$ can be quite difficult. However, for $L\rightarrow \infty$ the roots of (\ref{ZE}) are greatly simplified and the eigenvalues can be written as follows:
\begin{eqnarray}
\label{ev1}
\Lambda=4\pi_{r}\pi_{l} \;\;\mbox{for}\;\  -1 <  X , Y  < 3+2\sqrt{2}\\
\label{ev2}
\Lambda=-\pi_{r}\pi_{l}(\sqrt{Y}-\frac{1}{\sqrt{Y}})^2 \;\;\mbox{for}\;\;Y>3+2\sqrt{2}\;,\;-1< X < Y \\
\label{ev3}
\Lambda=\pi_{r}\pi_{l}(\sqrt{\vert X \vert }+\frac{1}{\sqrt{ \vert X \vert }})^2 \;\;\mbox{for}\;\;X<-1\;,\; Y>X
\end{eqnarray}
where $X$ and $Y$ correspond to $A$ and $B$ or $B$ and $A$. Now the largest relaxation time is given by $\tau=\vert Re[\ln \Lambda] \vert^{-1}$. If $A$ or $B$ are such that in the thermodynamic limit one of the roots of (\ref{ZE}) results in $\Lambda=1$, then the relaxation time diverges. Note that the eigenvalues (\ref{ev1})-(\ref{ev3}) do not always occur and their presence depends on whether the hopping probabilities $\pi_{l,r}$ and the reflecting probabilities $\pi_{r2,l2}$ are positive. \\  
Let us consider a simple example for which the roots of (\ref{ZE}) can be calculated exactly. For:
\begin{equation}
t_{34}=\frac{1-\beta}{\delta}\;\;,\;\;t_{24}=\frac{1-\gamma}{\alpha}
\end{equation}
we have:
\begin{equation}
\pi_{l2}=\pi_{l}^{2}\;\;,\;\;\pi_{r2}=\pi_{r}^{2}.
\end{equation}
In this case $A=B=0$ and the roots of (\ref{ZE}) are, according to (\ref{ev1}), phases and result in a relaxation time which in the large-system-size limit is given by :
\begin{equation}
\tau=\vert \ln (4\pi_{l}\pi_{r}) \vert^{-1}.
\end{equation}
At the phase transition point, where $\pi_{l}=\pi_{r}=\frac{1}{2}$ and the shock drift velocity is zero, the relaxation time is diffusive. 
\section{Concluding remarks}
In this paper we have introduced a family of driven diffusive systems on an open lattice under sublattice-parallel updating scheme. We have calculated the steady-state of the system using the matrix product approach with a two-dimensional matrix representation whose structure is quite  similar to that introduced in \cite{JM3} for the case when the steady-state of the system can be written in terms of a linear superposition of shocks with random walk dynamics. We have shown that the dynamics of a single shock in the system is similar to that of a single random walker which moves on a one-dimensional lattice with reflecting boundaries. We have calculated the hopping probabilities of the shock front in the bulk and also the probabilities of reflection from the boundaries. Using the random walk picture we have been able to calculate the relaxation times in the limit of large-system-size. \\
The asymmetric simple exclusion process is the only model which has already been introduced and studied under sublattice-parallel updating scheme. Apart from the model studied in this paper defined in (\ref{Rules1})-(\ref{Rules3}), we have also found two other processes containing only one species of particles under sublattice-parallel dynamics in which the random walker picture can be applied. In the first model the rules are:
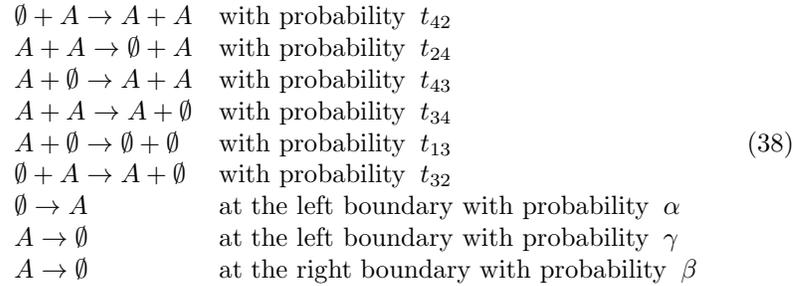
\begin{equation}
\label{MODEL1}
\begin{array}{ll}
\emptyset+A \rightarrow A+A & \mbox{with probability} \; \; t_{42} \\
A+A \rightarrow \emptyset+A & \mbox{with probability} \; \; t_{24} \\
A+\emptyset \rightarrow A+A & \mbox{with probability} \; \; t_{43} \\
A+A \rightarrow A+\emptyset & \mbox{with probability} \; \; t_{34} \\
A+\emptyset \rightarrow \emptyset+\emptyset & \mbox{with probability} \; \; t_{13}\\
\emptyset+A \rightarrow A+\emptyset & \mbox{with probability} \; \; t_{32} \\
\emptyset \rightarrow A  & \mbox{at the left boundary with probability} \; \; \alpha \\ 
A \rightarrow \emptyset  & \mbox{at the left boundary with probability} \; \; \gamma \\ 
A \rightarrow \emptyset  & \mbox{at the right boundary with probability} \; \; \beta
\\\end{array}
\end{equation}
provided that $\frac{t_{24}+t_{34}}{t_{24}+t_{34}+t_{42}}=\frac{t_{34}}{t_{34}-t_{32}+1}$. The rules for the second model are:
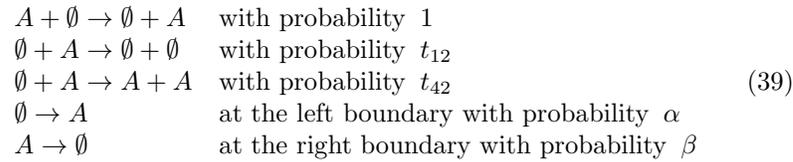
\begin{equation}
\label{MODEL2}
\begin{array}{ll}
A+\emptyset \rightarrow \emptyset+A & \mbox{with probability} \; \;1\\
\emptyset+A \rightarrow \emptyset+\emptyset & \mbox{with probability} \; \; t_{12} \\
\emptyset+A \rightarrow A+A & \mbox{with probability} \; \; t_{42} \\
\emptyset \rightarrow A  & \mbox{at the left boundary with probability} \; \; \alpha \\ 
A \rightarrow \emptyset  & \mbox{at the right boundary with probability} \; \; \beta
\\\end{array}
\end{equation}
without any constraints on the microscopic reaction probabilities. It can be shown that the steady-states of these models can be written as matrix product states and that the dynamics of an appropriately defined single shock front in these models is similar to that of a single random walker under parallel dynamics. The results will be published elsewhere; however, the classification of the two-state one-dimensional driven-diffusive models with open boundaries and nearest-neighbor interactions under parallel updating scheme with aforementioned properties, similar to \cite{KJS} for the models under random sequential updating scheme, remains an open problem. 
\ack
The authors would like to thank Gunter M. Sch\"utz for reading the manuscript and his enlightening comments.

\end{document}